\documentclass[prl,twocolumn,showpacs,floatfix]{revtex4}
\usepackage{graphicx}
\begin{document}
\title {Stability of the solutions of the Gross-Pitaevskii equation}
\author{A. D. Jackson$^1$, G. M. Kavoulakis$^2$, and E. Lundh$^3$}
\affiliation{$^1$Niels Bohr Institute, Blegdamsvej 17, DK-2100,
Copenhagen \O, Denmark \\
$^2$Mathematical Physics, Lund Institute of
Technology, P.O. Box 118, SE-22100 Lund, Sweden \\
$^3$Department of Physics, KTH, SE-10691, Stockholm, Sweden}
\date{\today}

\begin{abstract}

We examine the static and dynamic stability of the solutions of
the Gross-Pitaevskii equation and demonstrate the intimate
connection between them.  All salient features related to
dynamic stability are reflected systematically in static
properties.  We find, for example, the obvious result that
static stability always implies dynamic stability and present a
simple explanation of the fact that dynamic stability can exist
even in the presence of static instability.

\end{abstract}
\maketitle

\section{Introduction}

The question of excitations is central to the study of cold
atoms. Numerous experimental and theoretical investigations
have been devoted to the study of a variety of collective and
elementary excitations in these gases including vortex states,
solitary waves, and normal modes. When determining the
properties of an excited state, it is natural to consider two
kinds of possible instabilities -- static and dynamic.  In the
former case, one wishes to determine whether a state which
extremizes the energy is a genuine local minimum of the energy
subject to certain physically motivated constraints. Since the
wave function extremizes the energy, infinitesimal
perturbations will make no first-order change of the energy.
The state will be stable provided that an arbitrary
infinitesimal variation of the wave function necessarily
increases the energy to second order.

For the problem of dynamic stability, one conventionally
considers the temporal evolution of arbitrary infinitesimal
perturbations to the wave function using approximate linearized
equations (e.g., the Bogoliubov equations).  While the full
time-dependent equations conserve probability, the linearized
equations do not, and the resulting eigenvalue problem is
non-Hermitean.  If the corresponding eigenvalues are real, the
system is dynamically stable. In this case small-amplitude
time-dependent perturbations of the solution lead to bounded
motion about the original equilibrium position.  Instability is
indicated by complex eigenvalues, and the associated
exponential growth of small-amplitude perturbations drives the
system to a new state beyond the scope of the linearized
equations.

Here, we will explore the intimate connection between these two
seemingly-different criteria for stability.  Two main points
will emerge.  First, static stability implies dynamic
stability, but dynamic stability can exist even in the absence
of static stability.  Second, the transition from dynamic
stability to instability is reflected in the features of the
corresponding problem of static stability (subject to
appropriate constraints).  Here, inspired by the experiment of
Ref.\,\cite{Ketterle} and by the numerical studies of Refs.
\cite{Pu,Mottonen,Ohmi}, we will focus on the specific problem
of the stability (static and dynamic) of a doubly-quantized
vortex state.  Our study is, however, more general, and our
results can be applied to the study of the stability of any
problem described by the non-linear Gross-Pitaevskii equation.
One remarkable observation obtained from numerical simulations
of this problem \cite{Pu,Mottonen,Ohmi} is that the system
alternates between regions of dynamical stability and
instability as the strength of the interatomic interaction
increases. Since some (and probably all) regions of dynamical
stability coincide with regions of static instability, it is
useful to seek a simple description of this surprising
phenomenon.

In the following we first examine the questions of static and
dynamic stability separately. We will then demonstrate the
connection between them.

\section{Model}

We consider $N$ atoms subject to a spherically-symmetric
single-particle Hamiltonian, $h_i$, interacting through a
short-range effective interaction,
\begin{equation}
    H = \sum_{i=1}^N h_i + U_0 \sum_{i \neq j = 1}^N
    \delta({\bf r}_{i} - {\bf r}_{j})/2.
\label{intlab}
\end{equation}
Here $U_0 = 4 \pi \hbar^2 a/M$ is the strength of the effective
two-body interaction, $a$ is the scattering length for elastic
atom-atom collisions, and $M$ is the atomic mass.

For simplicity we assume strong confinement along the $z$ axis,
which is also chosen as the axis of rotation.  This assumption
implies that the cloud is in its lowest state of motion in the
$z$ direction, and the problem thus becomes two-dimensional
with higher degrees of freedom along the $z$ axis frozen out.
We also assume that $h_i$ is rotationally symmetric about the
$z$ axis (e.g., a harmonic oscillator Hamiltonian) with the
result that angular momentum is conserved.

\section{Energetic stability}

To be concrete, we will consider the problem of a
doubly-quantized vortex state.  We start by examining the
static stability of this state within the subspace of states of
the lowest Landau level.  We will generalize our arguments
below.  The corresponding nodeless eigenfunctions of the
(two-dimensional) single-particle Hamiltonian, $h_i$, will be
denoted as $\phi_m$, with $m$ the angular momentum; their
energy eigenvalues are $\epsilon_m$.

We now consider the state
\begin{eqnarray}
   \phi = c_0 \phi_0 + c_2 \phi_2 + c_4 \phi_4,
\label{defwf}
\end{eqnarray}
which describes a doubly-quantized vortex state when $c_2=1$
and $c_0 = c_4 =0$.  Since we wish to consider the effect of
small admixtures of states with $m=0$ and $m=4$, we expand the
energy to second order in $c_0$ and $c_4$ to obtain the
expectation value of the energy per particle
\begin{eqnarray}
    E = E_2 + c_0^* c_0 M_{00} + c_4^* c_4 M_{44} +
    M_{04} (c_0^* c_4^* + c_0 c_4),
\label{enstab}
\end{eqnarray}
where
\begin{eqnarray}
  E_2 &=& \epsilon_2 + \frac 1 2 g I_{2222}
\nonumber \\
  M_{00} &=& \epsilon_0 - \epsilon_2 + g (2 I_{0202} - I_{2222})
\nonumber \\
  M_{44} &=& \epsilon_4 - \epsilon_2 + g (2 I_{2424} - I_{2222})
\nonumber \\
  M_{04} &=&  g I_{0422} = g I_{2204},
\label{def}
\end{eqnarray}
with $I_{nmlk} \equiv \int \phi_n^* \phi_m^* \phi_l \phi_k \, d
{\bf r}$ and $g=N U_0$. The elements, $M_{mm'}$, are real due
to the Hermiticity of $H$.

The energy $E$ is then equal to
\begin{equation}
    E = E_2 + |c_0|^2 M_{00} + |c_4|^2 M_{44} + 2 |c_0| |c_4| M_{04}
 \cos(\theta_0 + \theta_4),
\label{enstaab}
\end{equation}
where $c_m = |c_m| e^{i \theta_m}$.  The absence of terms
linear in $c_0$ and $c_4$ is a consequence of the spherical
symmetry of $H$ and implies that $c_2 = 1$ is a solution to the
variational problem in this restricted space.  For fixed
$|c_0|$ and $|c_4|$, it is elementary that $E$ has extrema for
$\cos(\theta_0 + \theta_4) = \pm 1$.  In general, the linearly
independent curvatures of this quadratic energy surface are
obtained as the eigenvalues of the real symmetric matrix
\begin{eqnarray}
  M_s = \left( \begin{array}{cc}
M_{00} & M_{04}\\
M_{40} & M_{44}
\end{array} \right)
\end{eqnarray}
with $M_{40}=M_{04}^{\dag}$.

In this problem, it is evidently of physical interest to
consider the stability of the doubly quantized vortex subject
to the constraint that angular momentum is conserved. This
constraint will be satisfied if $|c_0| = |c_4| = |c|$, and the
energy $E$ then becomes
\begin{eqnarray}
    E = E_2 + 2 g |c|^2 (I_{0202} + I_{2424} - I_{2222} \pm I_{0422}).
\label{enstabb}
\end{eqnarray}
The doubly-quantized vortex will be an energy minimum if the
coefficient of $|c|^2$ is positive.  In other words, the
doubly-quantized vortex will be energetically stable if
\begin{equation}
 I_{0202} + I_{2424} - I_{2222} \ge I_{0422}.
\label{cond}
\end{equation}
This condition is not satisfied when only lowest Landau levels
are retained \cite{KMP}, and a doubly-quantized vortex state is
energetically unstable within this approximation.

Several comments are in order before turning to the dynamical
behavior of the same (truncated) problem.  We have tested the
stability of the state with respect to the admixture of states
$\phi_0$ and $\phi_4$.  A complete test of stability requires
the investigation of the admixture of $\phi_m$ for all $m$.
Fortunately, the spherical symmetry of $H$ ensures that the
generalized quadratic form of Eq.\,(\ref{enstab}) will only mix
states with $m=2 \pm k$.  With the inclusion of additional
values of $k$, the matrix $M_s$ becomes block diagonal and
leads to a sequence of equations for each $k$ completely
analogous to those considered above.  (The $k=2$ choice
considered here is known to be the most unstable case
\cite{KMP}.)  The problem of finding the eigenvalues of the
Hermitean matrix, $M_s$, is subject to familiar variational
arguments.  Expansion of the dimension of this matrix, either
by including more values of $k$ or by relaxing the restriction
to the lowest Landau level, can only decrease the smallest
eigenvalue of $M_s$.  Thus, the fact that a given state is
energetically unstable for a given choice of the finite space
used to construct $M_s$ is conclusive proof of instability.
Finally, we note that changing the state under investigation,
e.g., through the inclusion of higher Landau levels, leads to
fundamental changes in $M_s$.  Under such circumstances, simple
variational arguments cannot tell us whether the improved state
will be more or less stable.

\section{Dynamic stability}

As mentioned above, dynamic stability probes the temporal
behavior of the system.  We again start with the wave function
of Eq.\,(\ref{defwf}) and construct the Lagrangian to second
order in the small parameters $c_0$ and $c_4$. Variations with
respect to the coefficients $c_0^*$ and $c_4$ then lead to the
equations
\begin{eqnarray}
  i {\dot c}_0 &=& \epsilon_0 c_0 + 2 g c_0 c_2^* c_2 I_{0202} +
  g c_4^* c_2 c_2 I_{0422}
\nonumber \\
  -i {\dot c}_4^* &=& \epsilon_4 c_4^* + 2 g c_4^* c_2^* c_2 I_{2424} +
  g c_0 c_2^* c_2^* I_{0422}
\nonumber \\
  i {\dot c}_2 &=& \epsilon_2 c_2 + g c_2 c_2^* c_2 I_{2222},
\label{tm}
\end{eqnarray}
where the dot denotes a time derivative. We can solve these
equations with the ansatz $c_0(t)=c_0(0) {\rm e}^{-i(\mu +
\omega) t}$ and $c_4^*(t)=c_4^*(0) {\rm e}^{i(\mu - \omega)
t}$. (The time dependence of the solution, $\phi_2$, is given
by the factor ${\rm e}^{-i\mu t}$ with $\mu = \epsilon_2 + g
I_{2222}$.)  The resulting Bogoliubov equations can be written
in the form
\begin{eqnarray}
\left( \begin{array}{cc} \epsilon_0 - \mu + 2 g I_{0202} &
gI_{0422}
\\
-g I_{0422} & - \epsilon_4 + \mu - 2 g I_{2424}
\end{array} \right)
\left( \begin{array}{c}
c_0  \\
c_4^*
\end{array} \right) =
\nonumber \\
  \omega_d \left( \begin{array}{c}
c_0 \\
c_4^*
\end{array} \right).
\label{bog}
\end{eqnarray}
This equation can also be written as
\begin{equation}
M_s \,  \left( \begin{array}{c}
c_0 \\
c_4^*
\end{array} \right)= \omega_d \, \sigma
 \left( \begin{array}{c}
c_0 \\
c_4^*
\end{array} \right)\ \ {\rm with}\ \ \sigma \equiv
\left( \begin{array}{cr}
1 & 0\\
0 & -1\\ \end{array} \right),
\label{bog2}
\end{equation}
where $M_s$ is the matrix appearing in the static stability
problem.  The dynamic problem is thus governed by the
eigenvalues of the non-Hermitean Eq.\,(\ref{bog}), and these
eigenvalues can be complex.  Since $M_s$ is real Hermitean, the
dynamic eigenvalues are either real or come in conjugate pairs.
For each eigenfunction, $\psi_d$, with a complex eigenvalue,
$\omega_d$, $\psi_d^*$ will also be an eigenfunction with
eigenvalue $\omega_d^*$.  As a result, the roots move along the
real axis, touch (i.e., become degenerate) and then move off in
the complex plane.  Evidently, the existence of complex
eigenvalues indicates exponential divergence and dynamic
instability.

From Eqs.\,(\ref{bog}) or (\ref{bog2}) we see that the
Bogoliubov eigenvalues are real and the state under
investigation is dynamically stable provided that $I_{0202} +
I_{2424} - I_{2222} \ge I_{0422}$.  This condition is identical
to the condition for static stability found in
Eq.\,(\ref{cond}). As we show below in greater generality,
static stability always implies dynamic stability, but dynamic
stability can occur even in the presence of static instability.

\section{General results}

At this point it is useful to generalize our static and dynamic
formalisms to allow for the inclusion of an arbitrary number of
Landau levels in the description of both the doubly quantized
vortex and the potentially unstable states with $m=0$ and
$m=4$.  The matrix appropriate for the static problem now has
the Hermitean block form
\begin{eqnarray}
 M_s = \left( \begin{array}{cc}
 M_{00} & M_{04}  \\
 M_{40} & M_{44}
\end{array} \right),
\end{eqnarray}
with $M_s^{\dag} = M_s^T = M_s$.  Aside from its dimension, the
only change in the construction of $M_s$ lies in the
replacement of the lowest Landau state, $\phi_2$, by a
superposition of Landau states, $\psi_2$, which satisfies the
obvious Euler equation
\begin{equation}
   h_0 \psi_2 + g | \psi_2 |^2 \psi_2 = \mu \psi_2.
\label{euler}
\end{equation}
Here, $\mu$ is a Lagrangian multiplier introduced to ensure
that $\psi_2$ is normalized to unity.  It is understood that
the replacement $\phi_2 \to \psi_2$ is to be made as
appropriate in the integrals, $I_{nmlk}$, contributing to
$M_s$.  As before, positive eigenvalues of $M_s$ imply
energetic stability.

The dynamic Bogoliubov equations again assume the form of
Eq.\,(\ref{bog2}) and can be written as
\begin{eqnarray}
   M_s \left( \begin{array}{c}
c_0  \\
c_4^*
\end{array} \right) =
\omega_d \,  \sigma \left( \begin{array}{c}
c_0  \\
c_4^*
\end{array} \right),
\label{bog3}
\end{eqnarray}
where $M_s$ is again the real Hermitean matrix governing
energetic stability.  The quantities $c_0$ and $c_4$ now
represent column vectors; their dimensions need not be equal.
The reality of all eigenvalues, $\omega_d$, is again an
indication of dynamical stability.

As we have seen, $H$ is Hermitean and spherically symmetric.
The time evolution of an arbitrary wave function, $\psi$, is
governed by the full time-dependent Gross-Pitaevskii equation
\begin{equation}
h_0 \psi + g | \psi |^2 \psi = i \hbar \frac{\partial
\psi}{\partial t}.
\end{equation}
It thus comes as no surprise that energy and angular momentum
are, in general, constants of the motion.  On the other hand,
given that the (linearized) Bogoliubov equations do not even
conserve probability, it is not obvious that energy and
momentum are conserved when these equations are used to
describe the temporal evolution of the system.  This point
merits some attention. Consider a dynamic eigenvector of the
Bogoliubov equations, $\psi_d$, and its (possibly complex)
eigenvalue, $\omega_d$. Given the Hermiticity of $M_s$,
standard arguments reveal that
\begin{equation}
   (\omega_d - \omega_d^* ) \, \langle \psi_d | \sigma | \psi_d
\rangle = 0.
\label{conserve}
\end{equation}
This equation is evidently trivial when $\omega_d$ is real, but
it shows that $\langle \psi_d | \sigma | \psi_d \rangle = 0$
when $\omega_d$ is complex.  This tells us that the
probabilities of finding $m=0$ is equal to that for $m=4$ for
all times. In this sense, angular momentum is rigorously
conserved in spite of the approximations leading to the
Bogoliubov equations.

The extension of this familiar conservation law presumes that
the trial state $\psi_2$ is deformed by the inclusion (with
arbitrary amplitude) of a single Bogoliubov eigenvector with
complex eigenvalue. Angular momentum is not in general
conserved with arbitrary deformations of $\psi_2$ involving
either single Bogoliubov eigenvectors with real eigenvalues or
superpositions of Bogoliubov eigenvectors. Since the primary
rationale for studying the Bogoliubov equations at all is to
determine the existence or non-existence of complex
eigenvalues, this point is of interest in spite of such
caveats. Further, it suggests that we are most likely to reveal
relations between the problems of static and dynamic stability
if we consider the question of static stability subject to the
constraint of constant angular momentum. Indeed, this
constraint was imposed trivially in Eq.\,(\ref{enstabb}) above,
where it was crucial in establishing the identity of static and
dynamic stability criteria in the simple case of single Landau
levels.

For the more general problem considered here, it is easiest to
proceed by introducing real Lagrange multipliers, $\lambda$ and
$\eta$, in order to impose the constraints of overall
normalization and constant angular momentum, respectively, in
the static problem.  The constrained static eigenvalue problem
then reads
\begin{eqnarray}
   M_s {\bar \psi}_s =
\lambda \,  \left( \begin{array}{cc}
1 & 0\\
0 & \eta \end{array} \right) {\bar \psi}_s .
\label{omegas}
\end{eqnarray}
In practice, the real parameter $\eta$ is adjusted so that the
resulting static eigenfunctions ${\bar \psi}_s$ satisfy
$\langle {\bar \psi}_s | \sigma | {\bar \psi}_s \rangle = 0$.
The desired extrema of $\langle M_s \rangle$ then follow as
${\bar \omega}_s \equiv \langle {\bar \psi}_s | M_s | {\bar
\psi}_s \rangle$.  The similarity between the constrained
static problem of Eq.\,(\ref{omegas}) and the dynamic
eigenvalue problem of Eq.\,(\ref{bog3}) is now obvious. We
shall exploit this connection below.

\section{Connection between static and dynamic stability}

Given the fact that $\langle \psi_d | \sigma | \psi_d \rangle =
0$, we see that the inner product of $\langle \psi_d |$ with
Eq.\,(\ref{bog3}) also yields $\langle \psi_d | M_s | \psi_d
\rangle = 0$.  If $\psi_2$ is deformed by the inclusion (with
arbitrary amplitude) of a single Bogoliubov eigenvector with
complex eigenvalue, the energy is rigorously conserved and
identical to that obtained for the pure state $\psi_2$.  We can
use any $\psi_d$ with complex eigenvalue as a trial function to
provide a variational upper bound of zero for the lowest
constrained static eigenvalue, ${\bar \omega}_s$.  Said
somewhat more simply, it is elementary that the smallest ${\bar
\omega}_s$ cannot be positive if there exists a state $\psi_d$
such that $\langle \psi_d | M_s | \psi_d \rangle = 0$.
(Henceforth, we will consider only the constrained static
problem and will refer to it simply as the ``static problem''.)
Thus, if the Bogoliubov problem has complex eigenvalues, the
lowest eigenvalue of the static problem must be negative (or
zero). Conversely, if all ${\bar \omega}_s > 0$, none of the
Bogoliubov eigenvalues can be complex.  In other words, {\it
dynamic instability guarantees static instability, and static
stability guarantees dynamic stability}.  This establishes the
first rigorous connection between the two stability problems.
While this result may appear obvious, a similar argument shows
that it is equally impossible to find a state $\psi_d$ such
that $\langle \psi_d | M_s | \psi_d \rangle = 0$ if all
$\bar{\omega}_s < 0$. This yields the more surprising result
that {\it complete static instability also leads to dynamic
stability}.

In order to find additional ties between the static and dynamic
stability problems, it is useful to follow the trajectory of
two Bogoliubov eigenvalues (as a function of the coupling
constant) as they evolve from distinct real values to a
conjugate pair. To see this, it is sufficient to consider the
two-dimensional space of $(\psi_0, 0)$ and $(0, \psi_4)$.
(Here, $\psi_0$ and $\psi_4$ are of arbitrary dimension and
normalized.)  We thus write
\begin{eqnarray}
 M_s \approx \left( \begin{array}{cc}
 E_0 & g V  \\
 g V &  E_4
\end{array} \right),
\end{eqnarray}
with all quantities real.  The Bogoliubov problem of
Eq.\,(\ref{bog3}) is readily solved with the following results.
For $|g| < g_c = |(E_0 + E_4)/2V|$, the $\omega_d$ are real and
distinct.  The eigenvectors are real and do not satisfy the
constraint of constant angular momentum.  For $|g| = g_c$, the
eigenvalues are degenerate with $\omega_d = (E_0 - E_4)/2$. The
eigenvectors are identical and given as $( \psi_0 , - \psi_4
)/\sqrt{2}$. (Recall that the problem is non-Hermitean.) For $g
> g_c$, the eigenvalues form a conjugate pair.  The
eigenvectors also form a conjugate pair with a non-trivial
phase, and the magnitudes of their $\psi_0$ and $\psi_4$
components are equal (i.e., they conserve angular momentum).

We can also solve the static problem using Eq.\,(\ref{omegas})
in the same truncated basis.  The values of $\eta$ required to
impose the constraint of conserved angular momentum are found
to be $\eta = (E_4 \pm gV)/(E_0 + gV)$.  The corresponding
static eigenvalues are ${\bar \omega}_s = (E_0 + E_4)/2 \pm gV$
with ${\bar \psi}_s = (\psi_0 , \pm \psi_4) /\sqrt{2}$, as
expected. For $|g| < g_c$, the static eigenvalues are positive.
For $|g| > g_c$, one of the static eigenvalues is negative. For
$|g| = g_c$, one static eigenvalue is precisely zero with
${\bar \psi}_s = \psi_d = (\psi_0 , - \psi_4)/\sqrt{2}$. The
condition for a static eigenvalue of zero is the same as the
condition for degenerate dynamic eigenvalues.

We also see that the static and dynamic eigenvectors are
identical at this point.  This should come as no surprise.  At
$|g| = g_c$, the Lagrange multiplier, $\eta$, is $-1$. Equation
(\ref{omegas}), which determines ${\bar \psi}_s$, becomes
identical to the Bogoliubov equation, Eq.\,(\ref{bog3}), and
identical solutions must result. This result is general.  While
the form of these two equations is similar, the fact that $M_s$
is a real symmetric matrix forces ${\bar \psi}_s$ and ${\bar
\omega}_s$ to be real (up to a trivial phase).  When $\omega_d
\ne \omega_d^*$, it is clear that the associated dynamic
eigenvectors, $\psi_d$ and $\psi_d^*$ cannot be chosen equal.
Thus, the phase of $\psi_d$ is non-trivial, and $\psi_d$ cannot
be the desired solution to Eq.\,(\ref{omegas}). (The case of
$\omega_d$ real can be dismissed summarily, since the
amplitudes of $m=0$ and $4$ states are not equal.) When two
Bogoliubov roots are degenerate, however, $\psi_d$ satisfies
both the constraints of angular momentum conservation and
reality.  Under such conditions, ${\bar \psi}_s = \psi_d$, and
${\bar \omega}_s = 0$.  In other words, the number of conjugate
pairs of complex Bogoliubov eigenvalues changes by one when a
static eigenvalue passes through zero.

It might be thought that the number of conjugate pairs of
dynamic eigenvalues was equal to the number of negative ${\bar
\omega}_s$ and that a stronger statement was thus possible. As
we will show below, this is not the case.  The present
statement can nevertheless give us a useful corollary. Start
from a manifestly stable choice of $\psi_2$ for which all
${\bar \omega}_s > 0$ and dynamical stability is ensured. Vary
some external parameter (e.g., the coupling constant, $g$). As
demonstrated above, a single conjugate pair of Bogoliubov
eigenvalues will appear when the first ${\bar \omega}_s$ passes
through zero, and the system becomes dynamically unstable. With
every subsequent passage of an ${\bar \omega}_s$ through zero,
the number of conjugate pairs changes (i.e., increases or
decreases) by one.  It is then clear that an odd number of
${\bar \omega}_s$ must correspond to an odd number of conjugate
Bogoliubov pairs and, hence, dynamic instability.

\section{Dynamic stability in the absence of static stability}

As mentioned earlier, it is possible for the system to be
dynamically stable even in the presence of static instability.
This effect is more subtle but is also useful in tightening the
connection between the problems of static and dynamic
stability.  A simple calculation will be helpful.  Start with
two solutions to the constrained static problem of
Eq.\,(\ref{omegas}), $| {\bar 1} \rangle$ and $| {\bar 2}
\rangle$.  The corresponding static eigenvalues are ${\bar
\omega}_{s1}$, which is assumed to be negative, and ${\bar
\omega}_{s2}$, which can pass through zero.  Approximate the
Bogoliubov equations by truncating the basis to include these
two states and the related states $| \bar{3} \rangle = \sigma |
\bar{1} \rangle$ and $| \bar{4} \rangle = \sigma | \bar{2}
\rangle$.  (This calculation is exact when the dimension of
$M_s$ is equal to four.) Since these states are not necessarily
orthogonal, the Bogoliubov equations assume the form
\begin{equation}
   M_s \psi_d = \omega_d \Sigma \psi_d,
\label{nonorthog}
\end{equation}
with $(M_s)_{ij} = \langle \bar{i} | M_s | \bar{j} \rangle$ and
$(\Sigma)_{ij} = \langle \bar{i} | \sigma | \bar{j} \rangle$.
Both matrices are real symmetric. The first two diagonal
elements of $M_s$ are the constrained static eigenvalues,
$\bar{\omega}_{s1}$ and $\bar{\omega}_{s2}$.  The remaining two
diagonal elements, $\bar{\omega}_{s3}$ and $\bar{\omega}_{s4}$,
will be assumed to be positive.  The matrix, $\Sigma$, is
intimately related to the overlap matrix with elements $\langle
\bar{i} | \bar{j} \rangle$. Since the basis states conserve
angular momentum, the diagonal elements of $\Sigma$ vanish.
Given the origin of the $| \bar{1} \rangle$ and $| \bar{2}
\rangle$ as constrained extrema of $\langle M_s \rangle$, it is
clear that the elements of $M_s$ and $\Sigma$ are not
independent.  We shall ignore this fact in the following
qualitative arguments.

First, consider the case when these four states are maximally
linearly dependent.  State $| \bar{1} \rangle$ has been assumed
to be distinct from $| \bar{2} \rangle$ and is explicitly
orthogonal to $| \bar{3} \rangle$.  Let us assume it is
identical to $| \bar{4} \rangle$.  This immediately implies
that $| \bar{3} \rangle$ is identical to $| \bar{2} \rangle$.
The approximate Bogoliubov equation is thus reduced to a $2
\times 2$ matrix equation in the space of $| \bar{1} \rangle$
and $| \bar{2} \rangle$ and assumes the form
\begin{eqnarray}
   \frac{1}{\langle {\bar 1} | \sigma | {\bar 2} \rangle} \left(
\begin{array}{cc}
\langle {\bar 1} | M_s | {\bar 2} \rangle & {\bar \omega}_{s2}
\\
{\bar \omega}_{s1} & \langle {\bar 1} | M_s | {\bar 2} \rangle
\end{array}
\right) \, \psi_d = \omega_d \, \psi_d.
\nonumber \\
\label{invff}
\end{eqnarray}
The eigenvalues of this problem are
\begin{eqnarray}
   \omega_d = \frac 1 {\langle {\bar 1} | \sigma | {\bar 2}
\rangle}
 \left( \langle {\bar 1} | M_s | {\bar 2}
 \rangle \pm \sqrt{{\bar \omega}_{s1}
 \, {\bar \omega}_{s2}}  \right),
\end{eqnarray}
and the corresponding (unnormalized) eigenvectors are
\begin{equation}
   \psi_d \sim \sqrt{{\bar \omega}_{s2}} \, \, | {\bar 1} \rangle
+ \sqrt{{\bar \omega}_{s1}} \, \, | {\bar 2} \rangle.
\label{evecs}
\end{equation}
When ${\bar \omega}_{s2} > 0$, the Bogoliubov eigenvalues form
a conjugate pair. Further, Eq.\,(\ref{evecs}) shows that
$\psi_d$ acquires a non-trivial phase and that $\langle \psi_d
| \sigma | \psi_d \rangle \sim {\rm Re}{[({\bar \omega}_{s1}^*
{\bar \omega}_{s2})^{1/2}]}$ is zero.  The system is
dynamically unstable.  When ${\bar \omega}_{s2} = 0$, the
Bogoliubov eigenvalues are degenerate and real. In this case,
$\psi_d$ is equal to $| {\bar 2} \rangle$, which is real and
conserves angular momentum.  These results are all consistent
with those found above.  When ${\bar \omega}_{s2} < 0$,
however, the Bogoliubov eigenvalues are real and distinct. The
corresponding $\psi_d$ can now be chosen real, and angular
momentum is no longer conserved.  The presence of two negative
static minima of $\langle M_s \rangle$ is thus capable of
creating dynamic stability in spite of manifest static
instability.  This argument is actually more detailed than
necessary.  We have already seen that the number of conjugate
pairs of Bogoliubov eigenvalues necessarily changes by one
every time a static eigenvalue passes through zero. Linear
dependence has reduced the present problem to one spanned by a
two-dimensional space. The fact that ${\bar \omega}_{s1} < 0$
ensures that there are initially two complex eigenvalues in the
space.  We have seen in general that the number of complex
eigenvalues must change by two when ${\bar \omega}_{s2}$
crosses zero.  Since all eigenvalues of this $2 \times 2$
problem are already complex, the only possibility is thus that
the number of complex eigenvalues is reduced to zero with
dynamic stability as a consequence, as we have seen.  States $|
\bar{1} \rangle$ and $| \bar{2} \rangle$ are both essential to
this process.

Now consider the case where the $| \bar{i} \rangle$ are all
mutually orthogonal and maximally linearly independent.
Partition the matrix into $2 \times 2$ blocks spanned by the
states $| \bar{1} \rangle$ and $| \bar{3} \rangle$ and the
states $| \bar{2} \rangle$ and $| \bar{4} \rangle$,
respectively.  Each of the diagonal sub-blocks of $M_d$ now has
a form familiar from Eq.\,(\ref{invff}).  The diagonal elements
are equal in each case.  The products of off-diagonal elements
are $\bar{\omega}_{s1} \bar{\omega}_{s3}$ and
$\bar{\omega}_{s2} \bar{\omega}_{s4}$, respectively.  If the
off-diagonal blocks of this matrix are sufficiently small, the
Bogoliubov eigenvalues will be given by the eigenvalues of
these two $2 \times 2$ matrices.   Since $\bar{\omega}_{s1}
\bar{\omega}_{s3} < 0$ by assumption, this block produces the
original conjugate pair of dynamical eigenvalues independent of
the properties of $| \bar{2} \rangle$.  As $\bar{\omega}_{s2}$
passes through zero, $\bar{\omega}_{s2} \bar{\omega}_{s4}$
becomes negative, and an additional conjugate pair of
eigenvalues appears independent of the properties of $| \bar{1}
\rangle$. This behaviour will persist whenever the off-diagonal
blocks of $M_d$ are sufficiently small.

We thus see that there are two possible outcomes when a second
static extremum becomes negative.  If the corresponding states
are ``strongly'' coupled with $|\langle \bar{1} | \sigma |
\bar{2} \rangle| = | \langle \bar{1} | \bar{4} \rangle |
\approx 1$, the initial dynamic instability will be eliminated.
If these states are ``weakly'' coupled with $|\langle \bar{1} |
\sigma | \bar{2} \rangle| \approx 0$, a second unstable dynamic
mode will appear.  There is another and potentially fruitful
way to distinguish these alternatives. Consider the evolution
of the closed surfaces with $\langle M_s \rangle = 0$ which
bound domains within which $\langle M_s \rangle < 0$. [It is
assumed that $\langle M_s \rangle$ is explored with real trial
vectors subject to the constraint of constant angular momentum.
Hence, these vectors lie on a (hyper)torus.]  Every negative
extremum of $\langle M_s \rangle$ is evidently surrounded by
such a surface, and every such surface must contain at least
one negative extremum.  If there is one negative extremum,
there is one $\langle M_s \rangle = 0$ surface.  There are two
possible ways for the topology of these surfaces to change when
a second static extremum becomes negative.  A second $\langle
M_s \rangle = 0$ surface, not connected to the first, can
appear. In this case, the new negative extremum must be a local
minimum. Alternatively, the original surface can spread,
exploiting the fact that the manifold is periodic, and touch
itself.  The point of contact necessarily represents a new
extremum of $\langle M_s \rangle$, and a few moments thought
reveals that this extremum is necessarily a saddle point. (When
there are more than two negative static modes, independent
$\langle M_s \rangle = 0$ surfaces can merge in a similar
manner.)  Randomly drawn numerical examples suggest that the
appearance of a new $\langle M_s \rangle = 0$ surface
corresponds to the case of weak coupling discussed above, and a
new unstable dynamic mode emerges.  The merger of an $\langle
M_s \rangle = 0$ surface with itself corresponds to the case of
strong coupling, and the original unstable dynamic mode
disappears.  Although the evidence is admittedly slim, it is
tempting to guess more generally that each closed surface with
$\langle M_s \rangle = 0$ produces zero unstable dynamic modes
when it contains an even number of negative extrema of $\langle
M_s \rangle$ and one unstable dynamic mode when it contains an
odd number of such extrema.

As we have shown, the reality and Hermiticity of $M_s$ and the
non-Hermiticity introduced by the specific form of $\sigma$ are
not sufficient to provide a unique answer to the question of
dynamic stability in the presence of an even number of unstable
static modes.  (As noted above, an odd number of negative
static modes always implies dynamic instability.)  The question
of what constitutes ``strong'' or ``weak'' coupling would thus
seem to require more details regarding the physical system in
question.  For this reason, we now turn to the question of the
stability of vortex solutions to the physically relevant
Gross-Pitaevskii equation.

\section{Numerical results}

We now study the static and dynamic stability of a doubly
quantized vortex numerically. The single-particle Hamiltonian,
$h_0=-\hbar^2 \nabla^2 / (2 M) +M\omega^2 r^2/2$, represents a
harmonic oscillator potential of strength $\omega$.  In an
anharmonic potential, a doubly quantized vortex is
energetically stable for weak couplings, but in a purely
harmonic potential this is known not to be the case
\cite{lundh,jkl}.

\begin{figure}
\includegraphics[width=9cm,height=8cm]{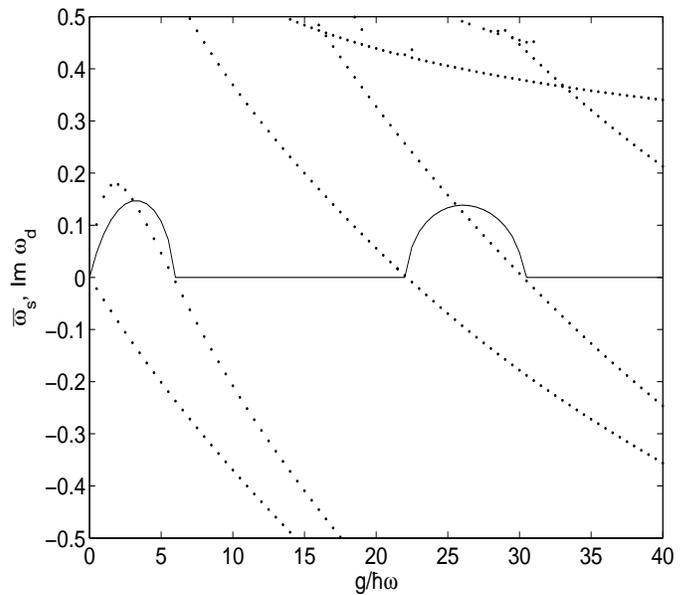}
\caption[]{Eigenvalues of the static and dynamic stability
matrices as functions of the coupling constant $g$, computed by
means of exact diagonalization in a truncated basis. Solid
lines represent the imaginary part of the eigenvalues of the
dynamic stability matrix, ${\rm Im} (\omega_d)$. Dots represent
the lowest eigenvalues of the static stability matrix,
$\bar\omega_s$. \label{fig:frequencies}}
\end{figure}

In our simulations Eq.\,(\ref{euler}) was first solved to find
the doubly quantized vortex.  This is equivalent to finding a
stationary solution to the Gross-Pitaevskii equation subject to
the constraint that the system has a 4$\pi$ phase singularity
at the origin.  This minimization was carried out within a
truncated basis composed of the $N_c$ lowest radially excited
states and is valid for small values of $g$ such that $g /(2
\hbar \omega) \lesssim N_c$. The static and dynamic eigenvalues
were then calculated according to Eqs.\,(\ref{bog3}) and
(\ref{omegas}) in a basis consisting of $N_x$ radially excited
states for both the $m=0$ and $m=4$ components. The computation
of the eigenvalues was carried out in Matlab. The results of
this calculation are shown in Fig.\,\ref{fig:frequencies}.
Results are shown for $N_c=N_x=8$, and we have checked
convergence with respect to both $N_c$ and $N_x$ for the range
of couplings displayed in the figure. The dynamic eigenvalues
$\omega_d$ coincide with those previously found numerically in
Refs.\,\cite{Pu,Mottonen,Ohmi}.

The figure clearly shows the correlation between the ``window''
structure of the complex frequencies and the eigenvalues,
$\bar\omega_s$, of the static problem in agreement with our
general arguments above. For small $g$ such that $g\ll
\hbar\omega$, the properties of the system follow from the
perturbative analysis of the lowest Landau level as described
above:  The system possesses one negative static eigenvalue and
therefore one pair of complex dynamic eigenvalues.  As the
coupling increases, a second static eigenvalue crosses zero at
$g/\hbar\omega\approx 6$, and the system becomes dynamically
stable despite its manifest static instability.  As a third
static eigenvalue crosses zero, one pair of dynamic eigenvalues
again become complex but become real again when a fourth static
eigenvalue becomes negative.  The numerical findings indicate
that as $g$ is further increased, a succession of such
alternations occurs with the result that the system is found to
be dynamically stable (unstable) when there are an even (odd)
number of negative static eigenvalues.

For the range of couplings, $g$, studied here, there is never
more than one complex pair of dynamic eigenvalues. It is also
seen that the system is always statically unstable. In the
limit of large $g$, well-known results for vortices in an
infinite system apply \cite{textbook}.  In that limit the
energy of two vortices increases monotonically with decreasing
separation, thereby implying that at least one eigenvalue of
the static stability matrix is negative.  Hence it is
reasonable to conclude that the system is statically unstable
for all values of $g$. We thus conclude that the ``window''
structure of alternating dynamically stable and unstable
regions arises from the non-trivial interplay between negative
modes of the static stability matrix.

\section{Conclusions}

The purpose of this paper has been to point out the intimate
connections between the static and dynamic stability of
solutions to the Gross-Pitaevskii equations.  We have shown
that (suitably constrained) static stability necessarily
implies dynamic stability, that the number of complex conjugate
pairs of dynamic eigenvalues changes by one every time a
constrained static eigenvalue passes through zero, and that an
odd number of negative static eigenvalues thus implies dynamic
instability.  Numerical investigations revealed that the
doubly-quantized vortex solution to the Gross-Pitaevskii
equation is statically unstable over the full range of coupling
constants explored but displays windows of dynamic stability.
The general nature of the arguments presented here suggests
that similar connections between the problems of static and
dynamic stability are likely to be wide-spread.  Further, we
believe that additional insight obtained by studying both
stability problems is likely to be well worth the minimal
additional effort required.

\section{Acknowledgments}
GMK acknowledges financial support from the European Community
project ULTRA-1D (NMP4-CT-2003-505457).

\end{document}